\documentclass[epj]{svjour}
\usepackage{epsfig} 
\usepackage{times}
\begin{document}
\title{Phenomenology of the ${\bf \Lambda/\Sigma^0}$ production ratio in ${\bf pp}$
collisions}
\author{A.~Sibirtsev\inst{1}, J.~Haidenbauer\inst{2}, 
H.-W. Hammer\inst{1}and U.-G.~Mei{\ss}ner\inst{1,2}} 
\institute{Helmholtz-Institut f\"ur Strahlen- und Kernphysik (Theorie), 
Universit\"at Bonn, Nu\ss allee 14-16, D-53115 Bonn, Germany \and
Institut f\"ur Kernphysik (Theorie), Forschungszentrum J\"ulich,
D-52425 J\"ulich, Germany}
\date{Received: date / Revised version: date}

\abstract{We show that the recently measured asymmetry in 
helicity-angle spectra of the $\Lambda$-hyperons produced in 
the reaction $pp{\to}K^+\Lambda{p}$ reaction and the 
energy dependence of the total $pp{\to}K^+\Lambda{p}$ cross 
section can be explained consistently by the same $\Lambda{p}$ 
final-state interaction. Assuming that there is no
final-state interaction in the $\Sigma^0{p}$ channel, as
suggested by the available data, we can also reproduce 
the energy dependence of the $\Lambda{/}\Sigma^0$ production 
ratio and, in particular, the rather large ratio observed 
near the reaction thresholds. 
The nominal ratio of the $\Lambda$ and $\Sigma^0$ production 
amplitudes squared, i.e. when disregarding the final-state 
interaction, 
turns out to be about 3, which is in line with hyperon production data 
from proton and nuclear targets available at high energies.
}


\authorrunning{A.~Sibirtsev et al.}
\titlerunning{Phenomenology of the ${\bf \Lambda/\Sigma^0}$ production ratio in 
${\bf pp}$ collisions}

\maketitle
One of the surprising results observed at the COSY accelerator facility 
is the large ratio of the $pp{\to}K^+\Lambda{p}$ to $pp{\to}K^+\Sigma^0{p}$ 
cross sections near the reaction thresholds~\cite{Sewerin,Kowina2}. 
This ratio is as large as 28~\cite{Sewerin} at very low energies and eventually 
approaches values around 3 with increasing energy. 
In the last few years several theoretical 
studies~\cite{Gasparian,Sibirtsev9,Laget,Shyam1,Shyam2,Dillig} 
appeared where different production scenarios were considered in order to
describe simultaneously both the $pp{\to}K^+\Lambda{p}$ and $pp{\to}K^+\Sigma^0{p}$ 
reaction cross sections and, thus, the $\Lambda{/}\Sigma^0$ ratio. 
These included $\pi$, $K$ as well as heavier meson exchanges, 
intermediate baryonic resonances coupled to the $K^+\Lambda$ and 
$K^+\Sigma^0$ channels as well as the $\Lambda{p}$ and $\Sigma^0{p}$ 
final-state interaction (FSI). But, as summarized in Refs.~\cite{Kowina2,Rozek1}, 
none of these models is able to reproduce the energy dependence of the 
$\Lambda{/}\Sigma^0$ ratio convincingly. 

A basic problem is that the relevance of the FSI on one hand 
side, and the interplay between the different baryonic resonances 
on the other side can not be resolved from considering only the total 
reaction cross section.
Therefore, in the present paper we follow a different strategy. 
We look not only at the total cross sections but also at
differential observables because some of them do allow to
distinguish between effects from the FSI and baryonic resonances,
as we pointed out recently \cite{Sibirtsev14}. Moreover, for the
reaction $pp{\to}K^+\Lambda{p}$ data for the relevant observables
are already available in the literature \cite{TOF}. Note that our 
study is completely phenomenological in the sense that we do not 
consider any specific reaction mechanism. But we take into account 
the effects from the $\Lambda{p}$ final-state interaction. 
We will argue that the differential data of Ref. \cite{TOF}
strongly support the interpretation of the measured
energy-dependence of the total $pp{\to}K^+\Lambda{p}$ cross
section in terms of FSI effects. At the same time, and in view
of the lack of any visible FSI effects in the reaction 
$pp{\to}K^+\Sigma^0{p}$ \cite{Kowina2}, we come to the conclusion that 
the energy-dependence of the $\Lambda{/}\Sigma^0$ cross section
ratio could likewise be governed primarily by FSI effects in the
$pp{\to}K^+\Lambda{p}$ channel. 
 
The role of the $\Lambda{p}$ FSI in
describing the $pp{\to}K^+\Lambda{p}$ cross section near
the reaction threshold was already addressed after the first 
experiments of the COSY-11 Collaboration~\cite{Balewski1,Balewski2} 
for excess energies below 7~MeV. But only further measurements
by the COSY-11 and COSY-TOF Collaborations at higher energies
explicitely indicated~\cite{Sewerin,Kowina2,Brandt} 
that the energy dependence of the $pp{\to}K^+\Lambda{p}$ cross 
section deviates from the phase space,  i.e. the expected $\epsilon^2$ 
behavior. (The excess energy $\epsilon$ is defined as
$\epsilon = \sqrt{s}{-}m_K{-}m_\Lambda{-}m_p$, where $s$ is the
squared invariant collision energy, while $m_K$, $m_\Lambda$ and $m_p$
are the masses of the kaon, the $\Lambda$-hyperon and the proton,
respectively.) 
This deviation from the phase-space behaviour -- in particular an 
enhancement at lower collision energies -- was interpreted
as an effect due to the $\Lambda{p}$ FSI in several investigations 
\cite{Gasparian,Sibirtsev9,Shyam1,Faldt,Hanhart}.

However, the near-threshold enhancement could also result from the
contributions of resonances coupled to the meson-baryon ($K^+\Lambda$) subsystem. 
Indeed, varying the mass and width of such resonances one can reproduce 
the energy dependence close to the reaction threshold too. 
For example, assuming that the $N^\ast$(1535) resonance
couples to the $K\Lambda$ system as strongly as to $\eta{N}$ the
authors of Ref.~\cite{Liu} came to the conclusion that the 
near-threshold enhancement observed in the $pp{\to}K^+\Lambda{p}$ 
cross section data is caused primarily by the $N^\ast$(1535) 
contribution. In their calculation the $\Lambda{p}$ FSI was entirely 
neglected! 
 
Thus, it is obvious that the 
low-energy enhancement of the $pp{\to}K^+\Lambda{p}$ cross
section can be described by the $\Lambda{p}$ FSI as well as by the
contribution of resonances coupled to the $K^+\Lambda$ system with
the resonance mass below $m_K{+}m_\Lambda$. 
As already mentioned, to resolve this issue one needs to analyze 
differential data. 
For example, it is obvious that a low-mass enhancement of 
the $\Lambda{p}$ invariant mass spectrum, $M_{\Lambda{p}}$, would 
reflect the importance of the FSI. Corresponding data on the 
$\Lambda{p}$ mass spectra were already available for some time, 
though only for $\epsilon{>}$250~MeV. Invariant mass spectra can 
be obtained from the $K^+$-meson momentum spectra measured at certain 
angles in $pp$ collisions~\cite{Hogan,Reed,Siebert}.
Although the $K^+$-meson spectra were
measured for the inclusive reaction, {\it i.e.} $pp{\to}K^+X$, the
contribution from the exclusive $pp{\to}K^+\Lambda{p}$ channel can be
well isolated by analyzing the missing mass spectra, $M_X$, below the
$\Sigma$-hyperon production threshold. The data provided in
Ref.~\cite{Reed,Siebert} at different proton beam energies and
$K^+$-meson production angles indicate a substantial enhancement at low
$\Lambda{p}$ invariant masses with respect to the pure phase space
distribution. Actually, the data from Ref.~\cite{Siebert} were even
used in attempts to determine the $\Lambda{p}$ effective-range
parameters~\cite{Gasparyan,Hinterberger}. Therefore, there is no doubt 
that the $\Lambda{p}$ FSI plays a substantial role in the reaction 
$pp{\to}K^+\Lambda{p}$. 

\begin{figure}[t]
\vspace*{-1mm}
\centerline{\hspace*{1mm}\psfig{file=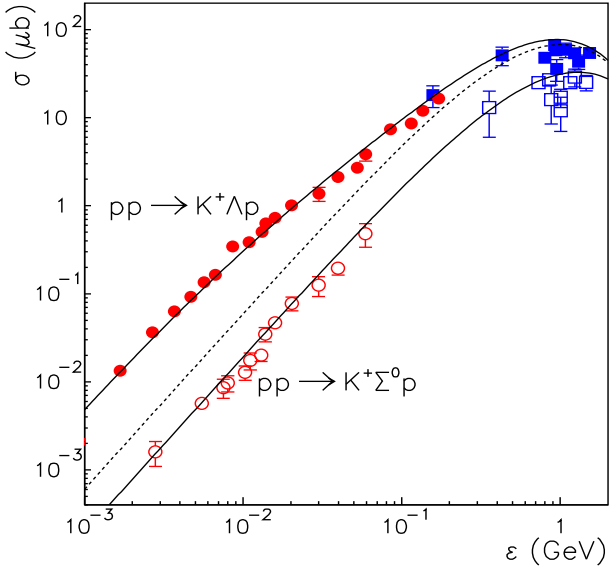,width=8.7cm,height=8.cm}}
\vspace*{-3mm}
\caption{Total cross sections for the $pp{\to}K^+\Lambda{p}$ (closed
symbols) and $pp{\to}K^+\Sigma^0{p}$ (open symbols) reactions as a
function of the excess energy $\epsilon$. Results from  
COSY~\cite{Sewerin,Kowina2,TOF,Balewski2,Brandt}
are indicated by circles, while the squares are data from 
Ref.~\cite{Baldini}. 
The solid lines are our results for the $\Lambda$ and $\Sigma^0$ reaction 
channels, respectively. The dashed line is obtained by switching
off the $\Lambda p$ final-state interaction. 
}
\label{kapro7a}
\end{figure}

On the other hand, it is still an open question how strongly the 
FSI contributes at COSY energies and, in particular, whether the 
$\Lambda{p}$ FSI is entirely responsible 
for the low-energy enhancement. To answer this question one 
requires differential observables for the $pp{\to}K^+\Lambda{p}$
reaction at COSY energies. Such data were recently provided
by the COSY-TOF Collaboration who measured~\cite{TOF}
the angular distributions
of the $\Lambda$-hyperon in reference to the proton direction in the
$K^+\Lambda$ center-of-mass system for the $pp{\to}K^+\Lambda{p}$
reaction at the beam momentum of 2.85 MeV/c ($\epsilon$ = 171 MeV). 
The angular spectra are
almost isotropic for the low invariant masses of the $K^+\Lambda$
system, {\it i.e.} at $M_{K\Lambda} < 1.69$ GeV. However, at larger
$K^+\Lambda$ masses the measured angular spectra exibit an anisotropy, 
which becomes more substantial  with increasing the $M_{K\Lambda}$. Note 
that the maximal $M_{K\Lambda}$ corresponds to the minimal invariant mass 
of the $\Lambda{p}$ system.

The measured angular spectra, which are related to the so-called 
helicity-angle ($\theta_H$) spectra, can be expressed~\cite{Sibirtsev14} in
terms of  the invariant masses of 
the $K^+\Lambda$ and $\Lambda{p}$ subsystems by expanding the
$\Lambda{p}$ invariant mass.
The helicity angle of the $\Lambda$-hyperon is then given as
\begin{eqnarray}  
\cos\theta_{H}{=}\left[2M_{K\Lambda}^2(m_p^2{+}m_\Lambda^2{-}M_{\Lambda{p}}^2)
{+}
(s{-}M_{K\Lambda}^2{-}m_p^2)\right.\nonumber \\
\left.\times (M_{K\Lambda}^2{+}m_\Lambda^2{-}m_K^2)\right]
\lambda^{-1/2}(s,M_{K\Lambda}^2,m_p^2)\nonumber \\
\times\lambda^{-1/2}(M_{K\Lambda}^2,m_\Lambda^2,m_K^2),
\label{helic}
\end{eqnarray}
where $\lambda(x,y,z){=}(x{-}y{-}z)^2{-}4yz$. For fixed $M_{K\Lambda}$
the angle $\cos\theta_H$=1 corresponds to the minimal invariant mass of
the $\Lambda{p}$ subsystem. Therefore, the $\Lambda{p}$ FSI should
manifest itself as an enhancement at forward $\theta_H$ angles. 
Eq.~(\ref{helic}) shows that applying different cuts on $M_{K\Lambda}$ one
can study the spectra at different $M_{\Lambda{p}}$ ranges and, at
sufficiently large $s$, one can isolate the $\Lambda{p}$ FSI.

In the following study we explore the effects of the $\Lambda{p}$ FSI.
It is included in our calculation within the 
Watson-Migdal approach \cite{WM} 
by factorizing the reaction amplitude in terms of a practically
constant production amplitude ${\cal M}_0$ and an FSI factor, 
i.e. 
\begin{eqnarray}
{\cal M} \approx {\cal M}_0 \times {\cal A}_{\Lambda p},
\label{jost0}
\end{eqnarray}
where the latter is taken to be the Jost function, 
\begin{eqnarray}
{\cal A}_{\Lambda p}(q)= \frac{q+i\beta}{q-i\alpha},
\label{jost1}
\end{eqnarray}
The parameters $\alpha$ and $\beta$ are related to the $\Lambda{p}$ 
effective-range parameters \cite{Sibirtsev14}. 
In the present work we employ the same values 
($\alpha$=-72.3 MeV/c, $\beta$=212.7 MeV/c corresponding to the 
$\Lambda{p}$ effective-range parameters $a = -1.8$ fm, $r = 2.8$ fm) 
that we already used in Ref.~\cite{Sibirtsev14}. With those 
the energy dependence of the $pp{\to}K^+\Lambda{p}$ reaction 
cross section could be described over a large energy range. 
For the production amplitude ${\cal M}_0$, however, we use here
a simple parametrization (see below) so that the total cross section 
can be evaluated analytically ~\cite{Sibirtsev14}:
\begin{eqnarray} 
\sigma (\epsilon){=}\frac{\sqrt{m_K m_N m_\Lambda}}{2^7 \pi^2 
(m_K{+}m_N{+}m_\Lambda)^{3/2}} \frac{\epsilon^2}{\sqrt{s^2{-}4sm_N^2}}
\, |{\cal M}_0|^2 \nonumber \\
\times\left[
1+\frac{4\beta^2-4\alpha^2}{(-\alpha+\sqrt{\alpha^2+2\mu\epsilon})^2}
\right] \ .
\label{cross1}
\end{eqnarray}
Here $\mu$ is the reduced mass of the hyperon-nucleon system. Note 
that the last term in the square brackets of Eq.~(\ref{cross1}) 
arises from taking into account the $\Lambda{p}$ FSI. 

With regard to ${\cal M}_0$ we allow for a smooth (exponential)
dependence on the excess energy in order to be able to connect
with the data at higher energies. To be concrete we use 
$|{\cal M}_0|^2{=}1.89\exp(-1.3\epsilon)$ $\mu$b ($\epsilon$
in GeV) which fits the $pp \to K^+\Lambda{p}$ cross section data 
\cite{Sewerin,Kowina2,TOF,Balewski2,Brandt,Baldini}
rather well as is shown in Fig.~\ref{kapro7a} by the solid line.
At the energy $\epsilon$=171 MeV where the helicity-angle 
distributions were measured 
we obtain 18.5 $\mu$b which is in reasonable agreement with the 
experimental value of $16.5{\pm}0.4$ $\mu$b \cite{TOF}. 
We would like to emphasize that the energy dependence introduced
by ${\cal M}_0$ is rather small over the range covered by
the COSY data (solid circles in Fig.~\ref{kapro7a}). It amounts to 
a variation of only about 20 \%. This has to be compared with the 
effect of the $\Lambda{p}$ FSI which produces an enhancement
of a factor 8 \cite{Sibirtsev14}, in the same energy range, 
cf. the solid and dashed curves in Fig.~\ref{kapro7a}.
 
\begin{figure}[t]
\vspace*{-1mm}
\centerline{\hspace*{1mm}\psfig{file=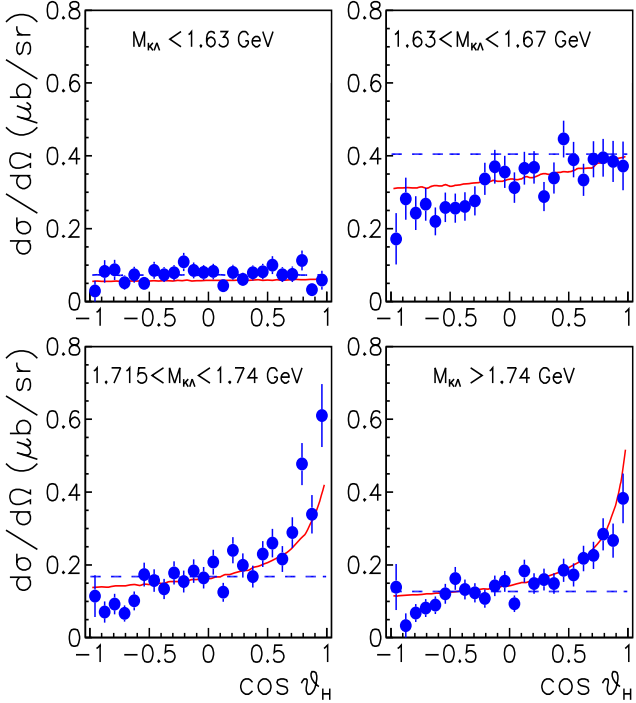,width=9.1cm,height=10.cm}}
\vspace*{-5mm}
\caption{The helicity-angle distribution of the $\Lambda$-hyperon for
different $K^+\Lambda$ invariant masses. The data are from
Ref.~\cite{TOF}, obtained at the excess energy $\epsilon$ = 171 MeV. 
The solid lines show our calculations including the $\Lambda{p}$ FSI, 
while the dashed are results obtained by switching off the FSI.}
\label{kasho3}
\end{figure}

Predictions for the helicity-angle distribution
of the $\Lambda$-hype\-ron are presented in Fig.~\ref{kasho3} 
together with the experimental angular spectra. 
Since the data are given in arbitrary 
units \cite{TOF} we have normalized them to our calculations. 
But we use an overall normalization for the experimental results, 
{\it i.e.} the same value for the different $M_{K\Lambda}$ cuts. 
Also
the results without the $\Lambda{p}$ FSI (dashed lines in 
Fig.~\ref{kasho3}) are based on the same normalization.

Obviously, 
the experimental result\footnote{In our convention the sign of 
$\theta_H$ is opposite to the one in the experimental
analysis~\cite{TOF}. In the following we use our sign convention since
it follows directly from the Dalitz plot
representation. Therefore, the experimental angular
distributions were transformed to helicity  $\theta_H$-angle spectra by
changing the sign of the experimental angle.} 
of Ref.~\cite{TOF} can be naturally explained in terms of the 
$\Lambda{p}$ FSI. The calculation without this FSI fails completely. 
Thus, we consider these angular spectra as a strong direct evidence for 
the presence of the $\Lambda{p}$ FSI. 
Most intriguing, however, is the fact that the specific strength
of the $\Lambda{p}$ FSI we employed reproduces the measured helicity-angle
distributions rather nicely and at the same time it
also yields an excellent description of the total $pp{\to}K^+\Lambda{p}$ 
cross section. This strongly suggests that the enhancement in the latter 
quantity at lower collision energies is primarily due to the $\Lambda{p}$ 
FSI. 
 
Let us now come to the reaction $pp{\to}K^+\Sigma^0{p}$. 
Corresponding cross section data are shown by
open symbols in Fig.~\ref{kapro7a}. Again, the circles are 
results of experiments at COSY~\cite{Sewerin,Kowina2}, while the squares
are from Ref.~\cite{Baldini}. The solid line shows
the calculation by Eq.~(\ref{cross1}) (replacing only $m_\Lambda$ 
by $m_\Sigma$) and without any $\Sigma^0p$ FSI, {\it i.e.} 
omitting the last term in the square brackets of Eq.~(\ref{cross1}). 
For the production amplitude ${\cal M}_0$ we use the same
form as before but with a readjusted normalization: 
$|{\cal M}_0|^2{=}0.61\exp(-1.3\epsilon)$ $\mu$b. 
We kept the smooth exponential energy dependence the same as 
before because (a) the $pp \to K^+\Sigma^0{p}$ reaction 
cross sections 
available for $\epsilon{>}$300 MeV have large uncertainties 
so that a determation of this parameter from a fit to those
data is not possible anyway, and 
(b) it is convenient to have the same energy dependence as
for the reaction $pp \to K^+\Lambda{p}$ because then 
the $\Lambda{/}\Sigma^0$ cross 
section ratio approaches a constant at high energies.

Obviously, the calculation without $\Sigma^0p$ FSI yields
already a perfect reproduction of the energy dependence 
of the $pp \to K^+\Sigma^0{p}$ data over the whole
considered energy range. Especially, in the COSY regime
the data are completely in line with the pure phase-space 
($\epsilon^2$) dependence. Based on this evidence one
would conclude that effects from the $\Sigma^0p$ FSI are 
much smaller than those in the $\Lambda p$ channel, as
already pointed out in Ref.~\cite{Kowina2}. 
Clearly, like for $pp{\to}K^+\Lambda{p}$ discussed above, 
one should keep in mind that firm conclusions can only 
by drawn once one has also inspected the helicity-angle
spectra for the $\Sigma^0$. Corresponding data at excess 
energies $60{<}\epsilon{<}210$ MeV have already been taken
by the TOF Collaboration and are presently analyzed~\cite{TOF1}.
It will be very interesting to see whether those data 
are in line with the absence of any $\Sigma^0{p}$ FSI as
conjectured from the total cross section. 
In such a case one expects that the $\cos\theta_H$ distribution is 
completely isotropic for large $K^+\Sigma^0$ invariant masses.
Anyway, the lack of apparent FSI effects in the $\Sigma^0{p}$ 
channel is certainly quite suprising. But it could be due to delicate
cancellations between effects resulting from the two possible 
transitions in the final state, namely  
$\Sigma^0{p} \to \Sigma^0{p}$ and $\Sigma^+{n} \to \Sigma^0{p}$.

\begin{figure}[b]
\vspace*{0mm}
\centerline{\hspace*{1mm}\psfig{file=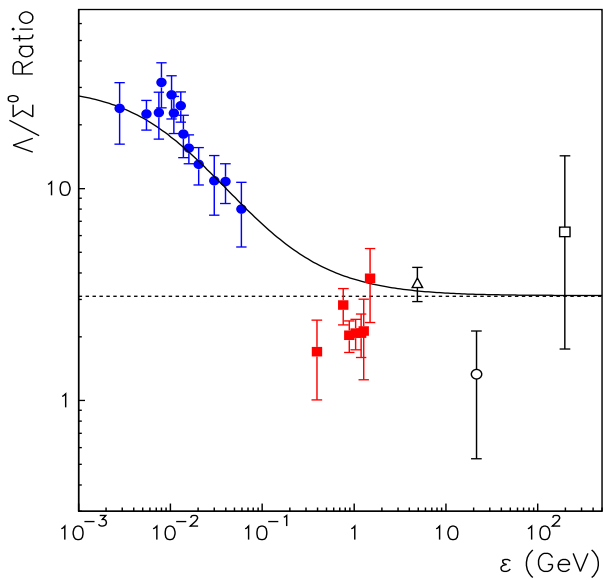,width=8.7cm,height=8cm}}
\vspace*{-2mm}
\caption{The $\Lambda{/}\Sigma^0$ cross section ratio as a function of the
excess energy $\epsilon$.  
The solid circles show the ratio obtained for the 
$pp{\to}K^+\Lambda{p}$ and $pp{\to}K^+\Sigma^0{p}$ reactions
at COSY~\cite{Kowina2}. Solid squares are $pp$ results from
Ref.~\cite{Baldini}.
The open triangle and open circle are ratios
measured in $p\,{\rm Be}$~\cite{Sullivan} and $p\,{\rm Ne}$~\cite{Yuldashev} 
collisions, respectively. The open square is the result from 
a $d\,{\rm Au}$ experiment~\cite{Buren}. The curves are cross section
ratios based on the $pp{\to}K^+\Lambda{p}$ results with $\Lambda{p}$ 
FSI (solid line) and without FSI (dashed line). 
} 
\label{kapro8}
\end{figure}

Based on our parametrizations of the production amplitudes
${\cal M}_0$ we can now calculate the ratio 
$\sigma_{(pp{\to}K^+\Lambda{p})}/\sigma_{(pp{\to}K^+\Sigma^0{p})}$.
The ratio of the squared production amplitudes alone yields
3.1, in rough agreement with the experimental ratio at higher
energies. Since the enhancement in the $pp{\to}K^+\Lambda{p}$ 
cross section at lower energies was found to be primarily due to
the $\Lambda{p}$ FSI, it follows immediately that also the 
near-threshold enhancement of the $\Lambda{/}\Sigma^0$ ratio 
must stem entirely from the $\Lambda{p}$ FSI. Our results are
illustrated in Fig.~\ref{kapro8}. The solid line is obtained with 
inclusion of the $\Lambda{p}$ FSI in Eq.~(\ref{cross1}) utilizing 
those FSI parameters ($\alpha$, $\beta$) that describe consistently
the helicity-angle distribution of the $\Lambda$-hyperon~\cite{TOF}. 
The dashed line in Fig.~\ref{kapro8} shows the ratio of the squared 
production amplitude for the $pp{\to}K^+\Lambda{p}$ and 
$pp{\to}K^+\Sigma^0{p}$ reactions. They are compared with
measurements from COSY~\cite{Kowina2} (solid circles) and 
with data at higher energies \cite{Baldini} (solid squares). 

It is worthwhile to mention that the $\Lambda{/}\Sigma^0$ ratio 
has also been discussed in the context of reactions with nuclear
targets. Quark model calculations predict~\cite{Anisovich,Bjorken} 
that in inclusive reactions at high energies the $\Lambda$-multiplicity
has to be much larger -- about 8 times -- than that of the 
$\Sigma$-hyperon. This does not violate $SU(3)$ symmetry but 
results from the decay of heavy baryonic resonances, which effectively 
enhances the $\Lambda$ production rate. 
Fig.~\ref{kapro8} contains the $\Lambda{/}\Sigma^0$ ratio observed in
$p\,{\rm Be}$ (open triangle)~\cite{Sullivan} and 
$p\,{\rm Ne}$ (open circle)~\cite{Yuldashev} collisions. The open
square is a very recent result~\cite{Buren} from the STAR Collaboration 
at RHIC obtained for $d\,{\rm Au}$ collisions at $\sqrt{s}$=200 GeV.

It is interesting to observe that the ratios for nuclear targets,
measured at high energies, are roughly in line with the results
from high-energy $pp$ collisions. Unfortunately, the new and still 
preliminary STAR result is afflicted by large uncertainties and, 
thus, precludes any firm conclusion concerning a possibly larger 
ratio with respect to that found in the $pp$ interactions.
Several authors have pointed out that the experimental ratio of
around 3 coincides with the ratio of the isospin multiplicity
of the $\Lambda$ and $\Sigma$'s \cite{Kowina2,Sullivan,Buren}. 
But we are not aware of any deeper reason why those two quantities
should be connected. 

In conclusion we analyzed the helicity-angle spectra of $\Lambda$
hyperons produced in the reaction $pp{\to}K^+\Lambda{p}$, measured 
recently by the COSY-TOF Collaboration. We argued that the observed 
anisotropy of the angular distribution at large invariant masses of 
the $K^+\Lambda$ system is a strong evidence for the presence of 
the $\Lambda{p}$ FSI. Adopting a specific strength of the $\Lambda{p}$ 
FSI the measured helicity-angle distributions could be reproduced 
quantitatively (within a Watson-Migdal approach). 
The same $\Lambda{p}$ FSI yields also an excellent description of the total 
$pp{\to}K^+\Lambda{p}$ cross section which strongly suggests that the 
enhancement in the latter quantity at lower collision energies is indeed
primarily due to the $\Lambda{p}$ FSI. Thus, our result casts 
considerable doubts on a recent interpretation of this enhancement 
in terms of a large contribution from the $N^*(1535)$ resonance \cite{Liu} 
(see also Ref.~\cite{Comment}).
In view of the lack of any visible FSI effects in the reaction $pp{\to}K^+\Sigma^0p$ 
this implies that the energy dependence of the $\Lambda{/}\Sigma^0$ ratio could
then be likewise driven by the $\Lambda{p}$ FSI. 
However, for the final conclusion on this issue data on the helicity-angle
distribution of the $\Sigma^0$-hyperon are necessary. 
The nominal ratio of the $\Lambda$ and $\Sigma^0$ production amplitudes squared, 
i.e. disregarding FSI effects, amounts to 3.1, which is in line with available 
data from proton and nuclear targets at high energies. 

\subsection*{Acknowledgements}
We would like to thank  M.~B\"uscher, D.~Grzonka, W.~Eyrich, 
J.~Ritman, \, E.~Roderburg, W.~Schroeder and
Yu. Valdau for useful discussions. 
This work was partially supported by the Deutsche
Forschungsgemeinschaft through funds provided to the SFB/TR 16
``Subnuclear Structure of Matter''. This research is part of the EU
Integrated Infrastructure Initiative Hadron Physics Project under
contract number RII3-CT-2004-506078. A.S. acknowledges support by the
COSY FFE grant 41760632 (COSY-085) and the JLab grant SURA-06-C0452.

\end{document}